\documentclass[prd,aps, preprintnumbers, showpacs,twocolumn, nofootinbib,superscriptaddress,notitlepage]{revtex4-1}
\usepackage{stmaryrd}
\usepackage[normalem]{ulem}
\usepackage{epsfig}
\usepackage{amsfonts}
\usepackage{amsmath}
\usepackage{slashed}
\usepackage{graphicx}
\usepackage{color}
\usepackage{mathtools}
\usepackage{subfigure}
\usepackage{graphicx}

\usepackage{amssymb,psfrag}
\usepackage{caption}
\usepackage{float}
\usepackage[table]{xcolor}
\usepackage{verbatim}

\newcommand{\beq}{\begin{eqnarray}}
	\newcommand{\eeq}{\end{eqnarray}}

\newcommand{\nn}{\nonumber}

\begin{document}
\title{Test for universality of short-range correlations in pion-induced Drell-Yan Process}
	
\author{Fei Huang}
\email{sps\_huangf@ujn.edu.cn}
\affiliation{ School of Physics and Technology, University of Jinan, 250022, Jinan, China,}	

\author{Shu-Man Hu}
\email{hushm2025@lzu.edu.cn}
\affiliation{Frontiers Science Center for Rare Isotopes, and School of Nuclear Science and Technology, Lanzhou University, Lanzhou 730000, China}
	
\author{De-Min Li}
\email{lidm@zzu.edu.cn}
\affiliation{School of Physics and Microelectronics, Zhengzhou University, Zhengzhou, Henan 450001, China}

\author{Ji Xu}
\email{xuji1991@sjtu.edu.cn}
\affiliation{Frontiers Science Center for Rare Isotopes, and School of Nuclear Science and Technology, Lanzhou University, Lanzhou 730000, China}

\date{\today}

\begin{abstract}
  We investigate nuclear modification and the universality of short-range correlation (SRC) in pion-induced Drell-Yan process. Employing nuclear parton distribution functions (nPDFs) and pion PDFs, the ratio of differential cross sections of different nuclei relative to the free nucleon is presented. A kind of universal modification function was proposed which would provide nontrivial tests of SRC universality on the platform of pion-induced Drell-Yan. This work improves our understanding of nuclear structure and strong interactions.
\end{abstract}
	
\maketitle		
	
\section{introduction }

In the Standard Model, most of the particles are composite. Investigating the internal structure of these composite particles has garnered considerable interest, and the parton distribution functions (PDFs) can help in this direction. The PDFs characterize the momentum distribution of partons within a composite particle such as $\pi, K, p$ and so forth. As the lightest of all hadrons, the $\pi$ meson acts as a mediating particle for the nuclear force, playing a vital role in binding nucleons together to form various atomic nuclei as well as in the creation of complex hadronic systems. Extensive theoretical research has been conducted based on the pion-induced Drell-Yan process, covering topics such as azimuthal spin asymmetries \cite{Gurjar:2024krn,Gurjar:2023uho}, EMC effect \cite{Wang:2018wfz,Chmaj:1985pp} and pion PDFs \cite{Nam:2012vm,Watanabe:2016lto,Hutauruk:2016sug,Lan:2019vui,Lan:2019rba,Watanabe:2019zny,Han:2018wsw,Chang:2014lva}.

Among the composite particles, the exploration of nucleons is also particularly prominent since they make up the atomic nuclei. The parton distributions of nucleons are called nuclear parton distribution functions (nPDFs). The EMC effect describes the modification of nPDFs in bound nuclei, first observed by the European Muon Collaboration in 1983 \cite{EuropeanMuon:1983wih} in muon-induced deep inelastic scattering (DIS) experiments, which found that the per-nucleon cross section for iron, relative to that of the deuteron, is reduced in the region where the Bjorken variable lies around 0.3 to 0.7. This finding diverges from initial expectations, which posited that the nucleon structure would not be affected by the distribution of the nuclear structure due to energy scale separation. Following that, the EMC effect has been validated in different nuclei \cite{EuropeanMuon:1988lbf,EuropeanMuon:1988tpw,NewMuonNMC:1990xyw,Seely:2009gt}.  Although many explanations have been proposed to explain the EMC effect, the mechanism of this effect is still unclear. Recently, research focusing on the relationship between the EMC effect and short range correlation pairs (SRCs) has attracted significant attention \cite{Hen:2016kwk,Weinstein:2010rt,CLAS:2005ola,Wang:2024ikx,Hen:2012fm,Hen:2014nza,CLAS:2018yvt,Xu:2019wso,Hatta:2019ocp,Wang:2024cpx}. SRC implies that nucleons are temporarily modified only when they fluctuate into an SRC pair,  these nucleon-nucleon pairs are characterized by high relative but low center-of-mass momentum in experiments. One of the key aspects of SRC is the universality, where the partonic structure from the correlated nucleon-nucleon pair is the same for all kinds of nuclei.
	
The lepton DIS probes the charge-weighted sum over all quark flavors and, in this sense, can be considered “flavor agnostic” \cite{Malace:2014uea}. Research exploring the EMC effect has encompassed many DIS experiments, including lepton-nucleus DIS \cite{Gomez:1993ri,BCDMS:1985dor,BCDMS:1987upi,Abrams:2024wgt} and neutrino-nucleus DIS \cite{WA25:1984ptx,Abramowicz:1984yk}. Besides various types of DIS experiments have been performed to investigate the EMC effect, it has also been verified in the time-like region using pion-induced Drell-Yan reactions in NA3 \cite{NA3:1981yaj} and NA10 \cite{NA10:1987hho} experiments. Theoretically, neutrino DIS has been utilized to explore the relationship between the EMC effect and the universality of SRC, one kind of universal modified function has been deduced within \cite{Yang:2023zmr,Huang:2021cac}. In addition, the pion-induced Drell-Yan process can also offer essential insights into the universality of SRC. In this paper, we will study the nuclear modification in pion-induced Drell-Yan, and the nuclei A are chosen to be $^{4}$He, $^{9}$Be, $^{12}$C, $^{27}$Al, $^{56}$Fe, $^{197}$Au, $^{208}$Pb. The numerical values of nPDFs are employed from EPPS21 parameterization \cite{Eskola:2021nhw}. For pion PDFs, we select the widely used pion PDFs parameterization which come from  xFitter group \cite{Novikov:2020snp}. A universal modification function will be presented which can be tested on future pion-induced Drell-Yan experiments.
	
This paper is organized as follows. In Sec.\,\ref{ratio  factor}, we give the differential cross-section ratio in Drell-Yan lepton-pair production in $\pi$-nucleus collisions. A ratio factor which to quantify the nuclear modification is also presented. The nPDFs are obtained from EPPS21 parameterizations  and pion PDFs come from  \texttt{xFitterPI\_NLO\_EIG} ($\pi^-$)  global fits. In Sec.\,\ref{modified ratio  factor}, the universal modification function constructed by the SRC scaling factor is proposed. Sec.\,\ref{summary} is reserved for conclusions.

\section{ratio  factor in Drell-Yan }
\label{ratio  factor}
	
In DIS, the measured structure functions are combinations of quark distributions. In principle, it is possible that theoretical models which yield almost identical results for EMC ratio in DIS can give different predictions for the Drell-Yan process \cite{Chmaj:1983jq,Chmaj:1984bu}. Moreover, the pion-induced Drell-Yan process could be pretty sensitive to the EMC effect \cite{Chang:2013opa}, making it an ideal platform for investigating the nuclear modification in nuclei.

In pion-induced Drell-Yan, a quark (antiquark) from the pion beam and antiquark (quark) from the nucleus annihilate via a virtual photon into a charged-lepton pair. The cross section of this process is \cite{Kulagin:2014vsa, Mitrofanov:2018roj}
\begin{eqnarray}\label{Drell-Yan}
  \frac{d \sigma_{\pi^- A}^2}{dx_\pi dx_2} &=& K \frac{4 \pi \alpha^2}{9s x_\pi x_2 } \sum_q e_q^2 \left[f_q^{\pi^-}(x_\pi, Q^2) f_{\bar{q}}^A(x_2, Q^2)  \right. \nonumber\\
  && \left.+ f_{\bar{q}}^{\pi^-}(x_\pi, Q^2) f_q^A(x_2, Q^2)\right]  \,.
\end{eqnarray}
where $\alpha$ is fine structure constant, $s$ is the center of mass energy squared, $x_2$ and $x_\pi$ is the Bjorken variable of the target nucleon and pion, respectively. $q$ is the flavor of quark, $f_{q,\bar{q}}^{\pi^- (A)}$ is quarks or antiquarks distributions in pion or the target nucleus. In Eq.\,(\ref{Drell-Yan}), the factor $K$ absorbs higher-order QCD corrections  as well as residual kinematical factors beyond the leading order. It can be expressed as \cite{Kenyon:1982tg, Conway:1989fs, Wijesooriya:2005ir}
\begin{eqnarray}\label{Kexpress}
  K \sim 1 + \frac{\alpha_s}{2\pi} \frac{4\pi^2}{3}+\mathcal{O}(\alpha_s^2) \,.
\end{eqnarray}
	
The quark and antiquark distributions in the nucleus can be constructed with the mass number $A$ and the proton number $Z$ in Eq.\,(\ref{Drell-Yan}),
\begin{eqnarray}\label{fA}
  f^{A}_q(x_2, Q^2) = \frac{Z f^{p/A}_q(x_2, Q^2)+(A-Z)f^{n/A}_q(x_2, Q^2)}{A} \,,
\end{eqnarray}
where $f^{p(n)/A}_q$ is quark nPDF of a proton (neutron) bound in the nuclei.

\begin{widetext}
Subsequently, the quark and antiquark distributions in pion in Eq.\,(\ref{Drell-Yan}) was integrated, since we are interested in the EMC effect of the nucleon,
\begin{eqnarray}
  \frac{d \sigma_{\pi^- A}}{dx_2} &=& K\frac{4 \pi \alpha^2}{9s  x_2} \sum_q e_q^2 \left[ f_{\bar{q}}^A(x_2, Q^2) \int_0^1 dx_{\pi} \frac{f_q^{\pi^-} (x_{\pi}, Q^2)}{x_{\pi}}+ f_q^A(x_2, Q^2) \int_0^1 dx_{\pi} \frac{f_{\bar{q}}^{\pi^-}(x_{\pi}, Q^2)}{x_{\pi}}  \right]  \,.
\end{eqnarray}
	
Furthermore, a ratio factor $R(d\sigma^{DY};x_2,Q^2)$ can be defined to quantify the nuclear modification,
\begin{eqnarray}\label{RDY}
  R(d\sigma^{DY};x_2,Q^2) &\equiv& \frac{d \sigma_{\pi^- A}/dx_2}{d \sigma_{\pi^- N}/dx_2}  \nonumber\\
  &=& \frac{\sum\limits_{q} e_q^2 \left[  f_{\bar{q}}^A(x_2, Q^2) \int_0^1 dx_{\pi} f_q^{\pi^-}(x_{\pi}, Q^2)/x_{\pi}  + f_q^A(x_2, Q^2) \int_0^1 dx_{\pi} f_{\bar{q}}^{\pi^-}(x_{\pi}, Q^2)/x_{\pi} \right] }{\sum\limits_{q} e_q^2 \left[  f_{\bar{q}}^N(x_2, Q^2) \int_0^1 dx_{\pi} f_q^{\pi^-}(x_{\pi}, Q^2)/x_{\pi}  + f_q^N(x_2, Q^2) \int_0^1 dx_{\pi} f_{\bar{q}}^{\pi^-}(x_{\pi}, Q^2)/x_{\pi} \right]}  \,,
\end{eqnarray}
here $f^N_{q(\bar{q})}$ in the denominator is the nPDF of a ``imaginary'' nucleus composed of one free proton and one free neutron. $f^N_{q(\bar{q})} = (f_{q(\bar{q})}^{p}+f_{q(\bar{q})}^{n}) / 2$, and the $f_{q(\bar{q})}^p$ denotes the quark (antiquark) PDF in a free proton and $f_{q(\bar{q})}^n$ is that of a free neutron. As a result of our analysis focusing solely on the ratios of differential cross sections between different target nuclei relative to the free nucleon, the contribution of $K$ factor would cancel out between the numerator and the denominator.

It is worth noting that the pion-induced Drell-Yan process involves the quark and antiquark distributions of both the pion and target nucleon. Since we are interested in the sub-structure of nucleon, we refer to the two terms in the numerator in Eq.\,(\ref{RDY}) as the contributions from the nucleon's antiquark and quark distributions, respectively. We depict these two contributions
\begin{eqnarray}
    &&\left\{
  \begin{aligned}
  & \sum\limits_{q} e_q^2 \left[ f_{\bar{q}}^A(x_2, Q^2) \int_0^1 dx_{\pi} f_q^{\pi^-}(x_{\pi}, Q^2)/x_{\pi}\right] \,,\\
  & \sum\limits_{q} e_q^2\left[f_q^A(x_2, Q^2) \int_0^1 dx_{\pi} f_{\bar{q}}^{\pi^-}(x_{\pi}, Q^2)/x_{\pi} \right] \,,
  \end{aligned}
  \right. \nn
\end{eqnarray}
in Fig.\,\ref{Drell-Yan} to compare their magnitudes where the nucleus here is selected as $^{56}$Fe. We use the result of $\pi^-$ PDF from \texttt{xFitterPI\_NLO\_EIG} global fits \cite{Novikov:2020snp}.  The $Q^2$ is fixed at $25\,\textrm{GeV}^2$, corresponding to the kinematic range applicable to most Drell-Yan experiments. The center of mass energy is $\sqrt{s}=21.8\,\textrm{GeV}$, which is in accordance with the beam energy of E615 experiments \cite{Conway:1989fs, Cerutti:2022lmb}.
The kinematic constraint $x_\pi x_2 = Q^2/s$ determines the lower integration limit $x_{\pi, \rm min} \approx 0.0526$ for $x_{2, \rm max} = 1$, hence the numerical integration is performed over $x_\pi \in [0.0526, 1]$.

From Fig.\,\ref{Drell-Yan}, it is apparent that within the EMC region of interest, $0.3<x_2<0.7$, the contribution from the nucleon's quark distributions is predominant. Therefore, the contribution from the nucleon’s antiquark distributions can be removed, leading to
\begin{eqnarray}
	R(d\sigma^{DY};x_2,Q^2) = \frac{a_{\bar{u}} f_u^A(x_2, Q^2) + a_{\bar{d}}  f_d^A(x_2, Q^2) }{ a_{\bar{u}} f_u^{N}(x_2, Q^2) + a_{\bar{d}} f_d^{N}(x_2, Q^2) }  \,.
\end{eqnarray}
To make the expressions above concise, we have adopted the following notations,
\begin{eqnarray}
  && a_q = e_q^2 \int_{0.0526}^1 dx_{\pi} \frac{f_q^{\pi^-}(x_{\pi}, Q^2)}{x_{\pi}}  \,, \nonumber\\
  && a_{\bar{q}} =  e_q^2 \int_{0.0526}^1 dx_{\pi} \frac{f_{\bar{q}}^{\pi^-}(x_{\pi}, Q^2)}{x_{\pi}} \,.
\end{eqnarray}

Making use of the quark distributions expressed in Eq.\,(\ref{fA}) and the isospin symmetry $f_{u(d)}^{n/A}=f_{d(u)}^{p/A},f_{\bar{u}(\bar{d})}^{n/A}=f_{\bar{d}(\bar{u})}^{p/A}$, the ratio factor $R(d\sigma^{DY};x_2,Q^2)$ can be rewritten as,
\begin{eqnarray}\label{RDY1}
  R(d\sigma^{DY};x_2,Q^2) &=& \frac{a_{\bar{u}} \frac{Z f^{p/A}_u(x_2, Q^2)+(A-Z)f^{p/A}_d(x_2, Q^2)}{A} + a_{\bar{d}}  \frac{Z f^{p/A}_d(x_2, Q^2) + (A-Z)f^{p/A}_u(x_2, Q^2)}{A} }{ a_{\bar{u}} \frac{ f^{p}_u(x_2, Q^2)+f^{p}_d(x_2, Q^2)}{2} + a_{\bar{d}} \frac{ f^{p}_d(x_2, Q^2)+f^{p}_u(x_2, Q^2)}{2} }  \,.
\end{eqnarray}
\end{widetext}

\begin{figure}[h]
  \centering
  \includegraphics[width=0.95\linewidth]{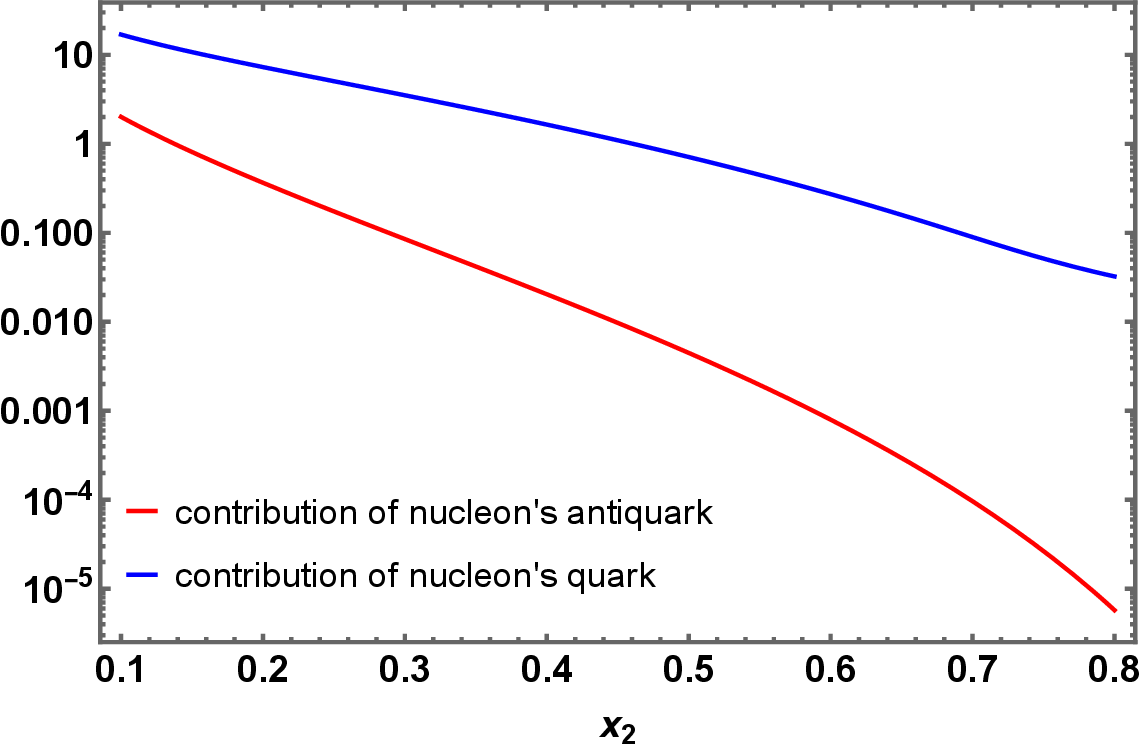}
  \caption{Comparison between the contributions from the nucleon's antiquark and quark distributions in Eq.\,(\ref{Drell-Yan}). The kinematics is chosen at $Q^2=25\,\textrm{GeV}^2$ and $\sqrt{s}=21.8\,\textrm{GeV}$.}
  \label{term12}
\end{figure}
	
\begin{figure}[htbp]
  \centering
  \label{R_xm}
  \includegraphics[width=0.95\linewidth]{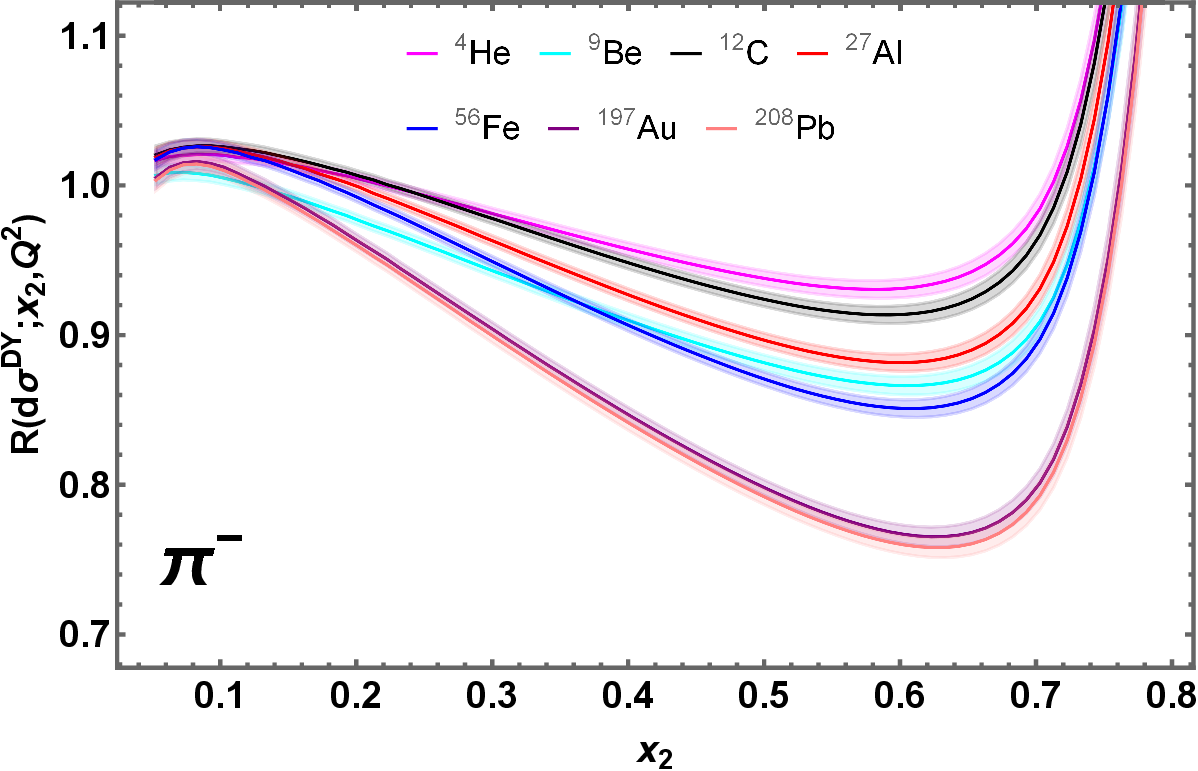}
  \caption{The ratio factors $R(d\sigma^{DY};x_2,Q^2)$ as a function of $x_2$ with $Q^2=25\,\textrm{GeV}^2$. The different color corresponds to different nuclei. }
  \label{R}
\end{figure}

To depict this ratio factor, we employ the quark nPDFs from EPPS21 and the pion PDFs from \texttt{xFitterPI\_NLO\_EIG} ($\pi^-$) in this work. The results are presented in Fig.\,\ref{R}. It can be seen clearly that within the region of $0.3\leq x_2 \leq 0.6$, this ratio factor decreases, covering a range roughly from 1.0 to 0.8, contingent upon the type of nucleus. The errors in this figure account for uncertainties in the quark nPDFs and the pion PDF. It is worth noting that the resulting errors are relatively small, due to our better understanding of these two kinds of distributions in the medium-$x$ region, as well as the cancellation of uncertainties between the numerator and denominator in the expression of the ratio factor $R$.

\section{the universality of SRC}
\label{modified ratio  factor}
	
The quark distribution of the proton  bound in nuclei $f^{p/A}_q$ can be defined by using the nuclear modification factor $R^{p/A}_q(x,Q^2)$ \cite{Eskola:2021nhw},
\begin{eqnarray}\label{fp}
  f^{p/A}_q(x,Q^2)=R^{p/A}_q(x,Q^2)f_q^p(x,Q^2) \,.
\end{eqnarray}
Inspired by the relationship between the EMC effect and SRC scaling factor, as well as the universality of SRC, we parameterize the modification of quark distributions of proton within bound nuclei \cite{CLAS:2019vsb}. We assume that quark PDFs of a proton bound in nuclei originate from both unmodified free protons as well as protons that belong to SRC pairs, with the assumption that all nuclear modifications stem from SRCs in EMC region,
\begin{eqnarray}\label{src_fp}
  f^{p/A}_q(x,Q^2)=\frac{1}{Z}\left[Zf^{p}_q(x,Q^2)+n^A_{SRC}\delta f^p_q(x,Q^2)\right]  \,,
\end{eqnarray}
here $n^A_{SRC}$ denotes the number of $np$ SRC pairs in nucleus A, and $\delta f^p_q(x,Q^2)$ reflects the difference of quark distribution between proton belonging to SRCs and free proton.

Through Eq.\,(\ref{fp}) and Eq.\,(\ref{src_fp}), one has
\begin{eqnarray}\label{universal equation}
  \frac{\delta f^p_q(x,Q^2)}{f^p_q(x,Q^2)}=\frac{R^{p/A}_q(x,Q^2)-1}{n^A_{SRC}/Z} \,.
\end{eqnarray}
The left-hand side of the above equation is nucleus-independent, implying that the right side of the equation is universal for all types of nuclei. Therefore, for two different nuclei A and B, we have
\begin{eqnarray}\label{universal equation 1}
  \frac{R^{p/A}_q(x,Q^2)-1}{n^A_{SRC}/Z_A}=\frac{R^{p/B}_q(x,Q^2)-1}{n^B_{SRC}/Z_B} \,.
\end{eqnarray}
	
Therefore, for the valance $u$ and $d$ quark, we can parametrize the distributions in A in terms of that in B,
\begin{eqnarray}\label{RuA RdA}
  R_{u_v(d_v)}^{p/A}(x,Q^2)=\frac{a_2^A}{a_2^B}\frac{A_A}{Z_A}\frac{Z_B}{A_B}\left(R_{u_v(d_v)}^{p/B}(x,Q^2)-1\right)+1 \,,\nonumber\\
\end{eqnarray}
where $a_2^A=(n^A_{SRC}/A)/(n^d_{SRC}/2)$ is the SRC scaling factor of nucleus A, it can be measured through the nuclear structure functions and its values are collected in \cite{Weinstein:2010rt,CLAS:2019vsb,Arrington:2012ax}.
	
We introduce the proton (neutron) excess constant $\epsilon_p^A$ ($\epsilon_n^A$) to address atomic nuclei with unequal numbers of protons and neutrons.
\begin{eqnarray}\label{correction facor}
  \epsilon_p^A=\frac{A/2-Z}{A}  \,, \quad
  \epsilon_n^A=\frac{A/2-(A-Z)}{A}  \,,
\end{eqnarray}
which can be concisely expressed as $\epsilon^A \equiv \epsilon_p^A = -\epsilon_n^A$. Substituting Eqs.\,(\ref{fp}) and (\ref{correction facor}) into Eq.\,(\ref{RDY1}), we can reexpress the ratio factor $R(d\sigma^{DY};x_2,Q^2)$ of nuclei A and B in terms of $\epsilon^A$,
\begin{widetext}
\begin{subequations}\label{epR}
\begin{eqnarray}
  && R(d\sigma^{DY}_A;x_2,Q^2) = \frac{\frac{1}{2}\Big[a_{\bar{u}}+a_{\bar{d}}\Big] \Big[R_u^{p/A}  f_u^p  \!+\! R_d^{p/A}  f_d^p \Big] +\epsilon^A\Big[a_{\bar{u}}-a_{\bar{d}}\Big]\Big[R_d^{p/A} f_d^p  \!-\! R_u^{p/A}  f_u^p\Big] }
  {\frac{1}{2}(f_u^p+f_d^p )\left[a_{\bar{u}}+a_{\bar{d}}\right]}  \,, \\\nonumber \\
  && R(d\sigma^{DY}_B;x_2,Q^2) = \frac{\frac{1}{2}\Big[a_{\bar{u}}+a_{\bar{d}}\Big] \Big[R_u^{p/B} f_u^p \!+\! R_d^{p/B} f_d^p\Big] +\epsilon^B \Big[a_{\bar{u}}-a_{\bar{d}}\Big] \Big[R_d^{p/B} f_d^p \!-\! R_u^{p/B} f_u^p\Big] }
  {\frac{1}{2}(f_u^p+f_d^p)\left[a_{\bar{u}}+a_{\bar{d}}\right]} \,. \\\nonumber
\end{eqnarray}	
\end{subequations}

Substituting Eq.\,(\ref{RuA RdA}) into Eq.\,(\ref{epR}a) and simplifying the result yields the following equation:
\begin{eqnarray}\label{RMAB}
  \frac{2Z_A}{A_A}\frac{R(d\sigma^{DY}_A;x_2,Q^2)-1}{a^A_2} +  \mathcal{O}(\epsilon^A) &=& \frac{2Z_B}{A_B}\frac{R(d\sigma^{DY}_B;x_2,Q^2)-1}{a^B_2} +  \mathcal{O}(\epsilon^B) \,.
\end{eqnarray}
\end{widetext}

\begin{figure}[htbp]
		\centering
		\label{RM_xm}
		\includegraphics[width=0.95\linewidth]{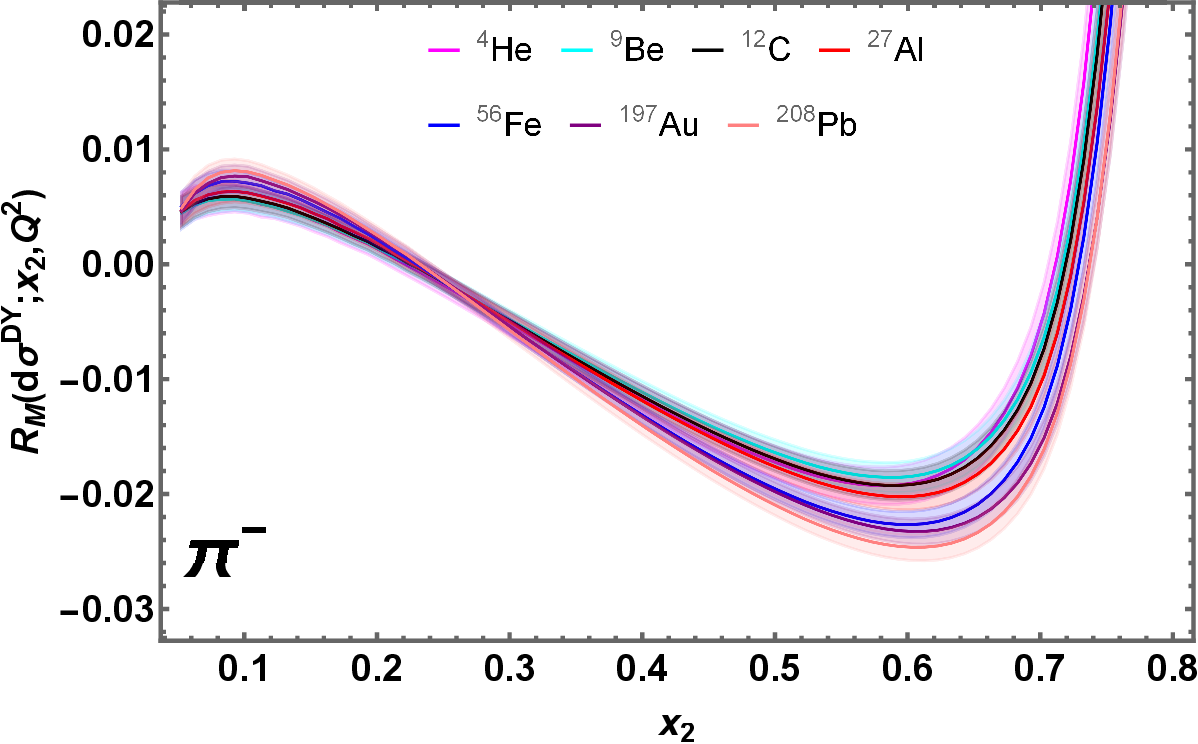}
		\caption{The shape of universal modification functions $R_M(d\sigma^{DY};x_2,Q^2)$ with $Q^2=25\,\textrm{GeV}^2$. The different color corresponds to different nuclei.}
		\label{RM}
	\end{figure}

This result suggests that the universality of SRC can be more precisely illustrated by introducing a kind of universal modification function which is experimentally measurable,
\begin{eqnarray}\label{RMDY}
  R_M(d\sigma^{DY};x_2,Q^2)=\frac{2Z_A}{A_A}\frac{R(d\sigma^{DY};x_2,Q^2)-1}{a^A_2} \,.
\end{eqnarray}	
In fact, one can consider the role of $R_M$ to be normalizing the ratio factor defined in Eq.\,(\ref{RDY}) by the SRC scaling factors. The $R_M$ has been plotted in Fig.\,\ref{RM}. One can observe that after this normalization, the ratios of different nuclei tend to shrink substantially, the range of these values have decreased from 1.0 to 0.8 to a range of 0.01 to -0.03. This conclusion remains valid within the uncertainties. This is a remarkable prediction, which validates the theoretical assumption presented in our paper, and indicates that the EMC effect is mainly caused by SRC in the region $0.3\leq x_2 \leq 0.7$. Fig.\,\ref{RM} proposes a nontrivial test of SRC universality in the pion-induced Drell-Yan process.
	
It is noteworthy that the term proportional to $\epsilon^{A(B)}$ in Eq.\,(\ref{epR}) violates the universal behavior of the modification factor in Eq.\,(\ref{RMDY}). However, as a higher-order correction, this term's contribution is approximately one order of magnitude smaller than the leading-order contribution. Therefore, we believe this correction term has little impact on the experimentally measurable $R_M$ in future studies.

SRC is one of the primary physics goals of the upcoming Electron-Ion Collider (EIC) experiment \cite{AbdulKhalek:2021gbh}. The universality presented here can be tested at the EIC. The upgrade to the COMPASS experiment, known as AMBER (Apparatus for Meson and Baryon Experimental Research), has also identified the SRC as a key focus of its research program \cite{Adams:2018pwt}. It will utilize pion beams colliding with nuclear targets to study the internal quark structure of nucleons. This setup makes the experiment suitable for investigating the proposed universality in this work.

\section{Summary}
\label{summary}

The EMC effect, which has been extensively studied over the past decades, indicates that the quark distributions in bound nuclei are modified compared to those in free nucleons. However, there is still no consensus for the cause of the EMC effect. Recently, the view that the EMC effect may be influenced by nucleon-nucleon short range correlation pairs has attracted a lot of attention. In addition to DIS, the pion-induced Drell-Yan experiment provides another avenue for the study of the relationship between the EMC effect and SRC, as well as the universality of SRC.

In this work, we studied the nuclear modification of different nuclei in the pion-induced Drell-Yan process. The ratio factors $R$ in Eq.\,(\ref{RDY}) are presented to quantify the EMC effect in different nuclei. These factors were found to significantly converge after the modification outlined in Eq.\,(\ref{RMDY}). From these results, one can come to the conclusion that the EMC effect can be described by the abundance of SRC pairs and the proposed modification functions are in fact universal. This work takes a small step forward in our understanding of nuclear structure and strong interactions.

\section{Acknowledgements}
This work is supported in part by National Natural Science Foundation of China under Grant No. 12505112, 12475098, and 12105247, Shandong Provincial Natural Science Foundation under Grant No. ZR2024QA138 and University of Jinan Disciplinary Cross-Convergence Construction Project 2024 (XKJC-202404).

	{}

\end{document}